# A Degradation Performance Model With Mixed-type Covariates and Latent Heterogeneity

Xuxue Sun[1], Wenjun Cai[2], Qiong Zhang[3], Mingyang Li[1]

[1]Department of Industrial and Management Systems Engineering, University of South Florida, USA

[2]Department of Materials Science and Engineering, Virginia Tech, USA

[3]School of Mathematical and Statistical Sciences, Clemson University, USA

**Abstract**

Successful modeling of degradation performance data is essential for accurate reliability assessment and failure predictions of highly reliable product units. The degradation performance measurements over time are highly heterogeneous. Such heterogeneity can be partially attributed to external factors, such as accelerated/environmental conditions, and can also be attributed to internal factors, such as material microstructure characteristics of product units. The latent heterogeneity due to the unobserved/unknown factors shared within each product unit may also exists and need to be considered as well. Existing degradation models often fail to consider (i) the influence of both external accelerated/environmental conditions and internal material information, (ii) the influence of unobserved/unknown factors within each unit. In this work, we propose a generic degradation performance modeling framework with mixed-type covariates and latent heterogeneity to account for both influences of observed internal and external factors as well as unobserved factors. Effective estimation algorithm is also developed to jointly quantify the influences of mixed-type covariates and individual latent heterogeneity, and also to examine the potential interaction between mixed-type covariates. Functional data analysis and data augmentation techniques are employed to address a series of estimation issues. A real case study is further provided to demonstrate the superior performance of the proposed approach over several alternative modeling approaches. Besides, the proposed degradation performance modeling framework also provides interpretable findings.

**Keywords:** Degradation performance data; Mixed-type covariates; Latent heterogeneity; Material microstructure characteristics

## 1. Introduction

Accurate modeling of degradation performance data is of great importance for achieving accurate reliability assessment and failure prediction nowadays for highly reliable product units with few and zero failure observations. Due to the varied product characteristics and the influences of many factors throughout the product design, manufacturing and operational phases, product units often exhibit highly heterogeneous degradation performance over time. To account for such degradation performance heterogeneity and to improve reliability assessment accuracy, many of the existing degradation models focused on modeling degradation performance data with covariates by incorporating various external influencing factors and quantifying their influences [1, 2]. At design phase, by investigating external factors, such as design settings or accelerated conditions (e.g., accelerated voltage, load, temperature), reliability engineers are able to either identify the most appropriate design changes for reliability improvement or achieve cost-effective reliability evaluation via analysis of accelerated degradation testing data. At operational phase, by investigating external factors, such as field operating conditions (e.g., usage rate, field temperature), reliability engineers are able to implement more accurate failure prognosis and initialize cost-effective condition-based maintenance actions. However, there is limited research to further extract and incorporate reliability relevant material characteristics of product units as internal factors and integrate them with external factors (e.g., accelerated/environmental conditions) for improving prediction accuracy of degradation performance outputs.

With the advancement of sensing technologies and material property characterization techniques, such as scanning electron microscope and transmission electron microscope [3], rich material information, such as microstructure characteristics, of product units become readily available or can be easily accessible. To extract rich characteristics information from material microstructure images, engineers often utilize informative material descriptors, such as two-point correlation function, radial distribution function and lineal-path function [4], and many of them are represented in the functional form rather than the scalar form. For instance, the two-point correlation is a functional feature curve over spatial distance to describe the spatial heterogeneity of material microstructure at microscopic scale, which often reflects the reliability-related product properties at macroscopic level (e.g., strength, hardness). Incorporating such functional covariates as internal reliability influencing factors and integrating them together with other external factors has a great opportunity for improving degradation performance modeling accuracy as compared to the existing models which only consider external factors. It will also help to improve the understanding of how material characteristics will influence the degradation performance of the product and further identify feasible material processing strategies to modify material settings for degradation-induced failure mitigation and reliability improvement. After considering the influences of both the internal and external observed factors, it is still possible that the degradation performance within each product unit may be highly correlated due to the latent heterogeneity. Such latent heterogeneity is essentially caused by the influence of many unobserved/unknown factors shared within each unit. Thus, there is a need to develop a generic degradation performance modeling framework which can simultaneously incorporate mixed-type covariates (e.g., both functional covariates of internal material microstructures and scalar covariates of external accelerated/environmental conditions) and latent heterogeneity to improve the degradation performance modeling accuracy.

In the existing reliability data modeling literature, different statistical/stochastic models have been developed to analyze different types of reliability data with covariates, such as failure counts data with covariates [5, 6] and time-to-failure data with covariates [7-9]. They mainly considered the scalar covariates that represent the external influencing factors, such as voltage, load, temperature and humidity, while they failed to incorporate internal factors, such as material characteristics information. In addition to the above reliability data types, degradation data is another important type of reliability data. For degradation data modeling in general, different data-driven models have been developed, such as continuous stochastic process [10-13], Markov-based models [14, 15], and general path models [16, 17]. Many of these modeling approaches mainly focused on characterizing the heterogeneity of degradation performance outputs as a whole without explicitly incorporating covariates as additional inputs to explain part of the heterogeneity. To characterize degradation performance heterogeneity with covariates, existing degradation models often considered scalar covariates that represent external factors, such as environmental conditions [18-20]. There is limited recent studies which account for the influences of material characteristics on degradation performance output. Park et. al. [21] incorporated a scalar covariate into degradation performance modeling which represented the aggregate-level material information. Si et. al. [22] considered functional covariate in their degradation model and incorporated detailed material microstructure information. However, these approaches failed to jointly consider both the mixed-type (i.e., functional and scalar) covariates and their potential interaction. The latent heterogeneity caused by the unobserved factors within each product unit was not addressed in these models as well.

To address the above research gaps, we propose a generic statistical degradation performance modeling framework to account for both the observed mixed-type covariates and the latent heterogeneity. The mixed-type covariates consist of (i) the functional covariates which capture the internal material microstructure characteristics of product units, and (ii) the scalar covariates which capture the external environmental conditions elevated in the context of accelerated degradation test. Moreover, a model estimation algorithm is developed to jointly quantify the influences of mixed-type covariates and the latent heterogeneity, and further to examine the potential interaction between functional and scalar covariates. Functional data analysis and data augmentation techniques are employed to address a series of estimation challenges, such as the infinite dimensionality of functional covariates and joint estimation of observed factors' effects and latent variable. To demonstrate the effectiveness of the proposed approach, we present a real case study using accelerated tribological degradation data of test units of copper alloys and demonstrate the superior performance of proposed model over several alternative degradation performance modeling approaches.

The rest of this paper is organized as follows. Section 2 describes the formulation of proposed degradation modeling framework and introduces the concepts of functional material descriptor, followed by a detailed elaboration of the developed algorithm of model estimation. Section 3 presents a real case study to illustrate the proposed work and further demonstrate its outperformance (e.g., accurate prediction performance, appealing model interpretation) over several alternative models. Section 4 draws the conclusive remark of this paper.

## 2. Methodology

To capture the influences of both environmental conditions and material characteristics as well as the latent effects of unobserved factors, we propose a generic degradation performance modeling framework with mixed-type covariates and latent heterogeneity, as shown in Figure 1. For each test unit, the degradation performance outputs are collected in the accelerated degradation test and are used as responses to develop the degradation performance model. Besides, the material microstructure information of each test unit is collected via advanced sensing technology. With the obtained microstructure images, we employ statistical measures to extract the functional features of material characteristics. We then use scalar covariates to represent external environmental conditions (e.g., temperature, loads and humidity) and utilize functional covariate to represent internal material characteristics. The proposed framework incorporates both mixed-type (e.g., scalar and functional) covariates and accounts for their potential interaction as well as latent heterogeneity. Further, we employ finite basis approximation technique to address the infinity dimensionality issue of functional covariates and apply data augmentation technique with expectation maximization estimation method to jointly estimate both mixed-type covariates and latent variable. The proposed modeling framework can improve prediction accuracy via considering both influences of mixed-type covariates and latent heterogeneity. In addition, the proposed framework is able to identify important influencing factors and quantify their effects on product reliability performance (e.g., accelerating/decelerating degradation process). With the quantified effects of observed influencing factors, the proposed framework can facilitate optimal test design and further improve product reliability. Besides, the proposed framework can also quantify the latent heterogeneity of individual test unit and help with future data collection. The details will be elaborated in the following subsections.

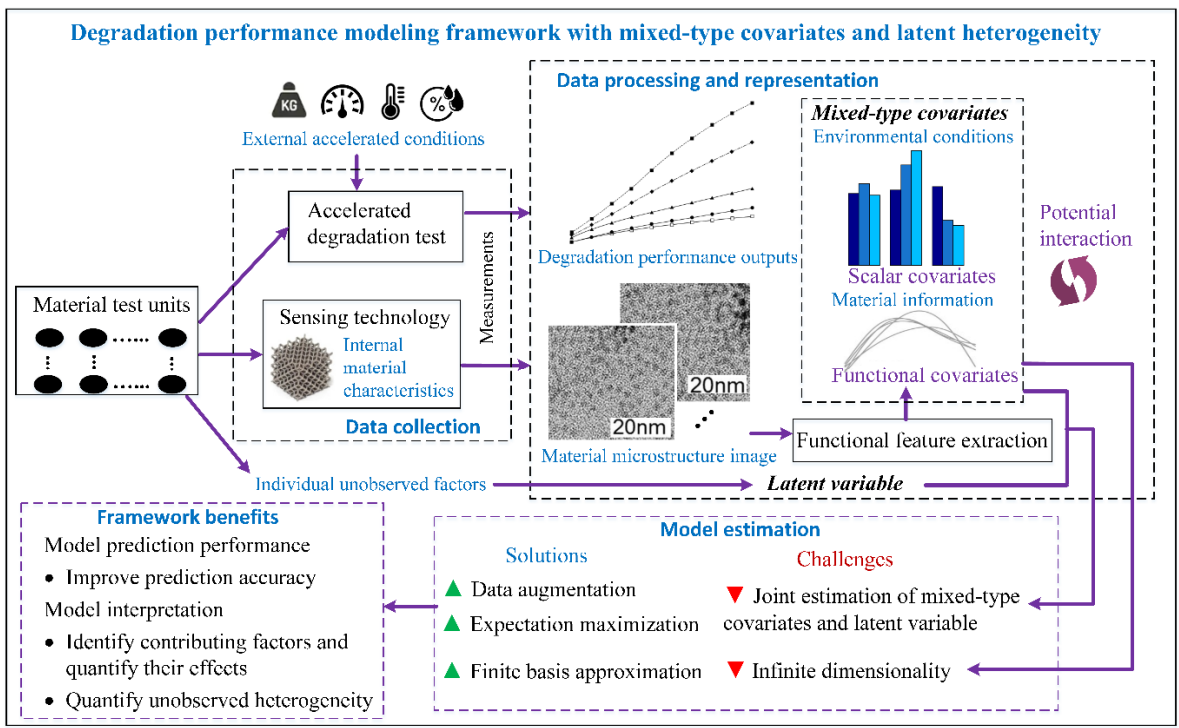

Figure 1: Overview of proposed degradation performance modeling framework

## 2.1. Model Formulation

Considering a population of $N$ test units, the observed degradation performance of test unit $i$ at time $t_{ij}$ is donated as $y_{ij}, \forall i = 1, ..., N, j = 1, ..., m_i$ where $m_i$ is total number of degradation performance measurements of test unit $i$. The degradation performance of each test unit may be influenced by both external factors, such as accelerated environmental conditions, and internal factors, such as the material characteristics. The former can often be characterized by scalar covariates to reflect a specific stress level of the accelerated environmental condition, while the latter can often be characterized by functional material descriptor due to their rich representation of material microstructure characteristics. In addition to the above influences of observed factors, it is still possible that each test unit may be influenced by a set of unobserved/unknown factors shared within each test unit, which causes the correlation among repeated degradation performance measurements of single test unit over time. To simultaneously quantify both the influences of the observed mixed-type (i.e., scalar and functional) covariates, their potential interaction effect and the influence of unobserved/unknown factors shared within each test unit, the proposed bi-level degradation performance model can be generically formulated as follows

Response level:
$$y_{ij} = g_i(t_{ij}, \theta_i, \Psi) + \epsilon_{ij} \approx \sum_{l=0}^{L} \eta_{li}\phi_l(t_{ij}) + \epsilon_{ij}, \quad i = 1, ..., N, j = 1, ..., m_i \quad (1a)$$

Coefficient level:
$$\eta_{li} = \nu_l + \boldsymbol{\beta}_l^T \boldsymbol{x}_i + \sum_{s=1}^{S} \int_{\mathbb{R}_s} \alpha_{ls}(r) Z_{is}(r) dr$$
$$+ \sum_{p=1}^{P} x_{ip} \left( \sum_{s=1}^{S} \int_{\mathbb{R}_s} \rho_{lps}(r) Z_{is}(r) dr \right) + \gamma_{li}, \quad i = 1, ..., N, l = 0, ..., L \quad (1b)$$

At response level, the degradation performance outputs over time of test unit $i$ can be captured by a unit-specific nonlinear function $g_i(\cdot)$ and an error term $\varepsilon_{ji} \sim N(0, \sigma_\varepsilon^2)$ where $\sigma_\varepsilon^2$ is the variance of measurement error. In the unit-specific nonlinear function mapping, $\Psi$ is a vector of fixed effect parameters and $\theta_i$ is a vector of random effect parameters of test unit $i$. To improve the model interpretation and estimation tractability, the nonlinear function $g_i(\cdot)$ can be further approximated by a set of basis functions $\{\phi_l(\cdot), \forall l = 0, ..., L\}$ and unit-specific basis coefficients $\{\eta_{li}, \forall l = 0, ..., L\}$. $\phi_0(\cdot) = 1$ and $\eta_{0i}$ is the grand mean function. Different basis functions, such as polynomial basis and spline basis, can be considered to capture the nonlinear curvature of degradation performance over time while unit-specific basis coefficients capture the individual heterogeneity of degradation performance. To further capture such individual heterogeneity, we decompose each coefficient $\eta_{li}$ into five components at coefficient level, namely, (i) population-level component $\nu_l$, which captures the population average degradation pattern at $l^{\text{th}}$ coefficient level among all test units; (ii) the individual heterogeneity at $l^{\text{th}}$ coefficient level explained by the marginal effect of observed scalar covariates $\boldsymbol{x}_i = [x_{i1}, ..., x_{iP}]^T$ and covariates coefficients $\boldsymbol{\beta}_l = [\beta_{l1}, ..., \beta_{lP}]^T$; (iii) the individual heterogeneity at $l^{\text{th}}$ coefficient level explained by the marginal effect of observed functional covariates $Z_{is}(r)$ with support space $\mathbb{R}_s$ and covariates coefficients functions $\alpha_{ls}(r), \forall s = 1, ..., S$; (iv) the individual heterogeneity at $l^{\text{th}}$ coefficient level explained by the interaction effect between observed scalar and functional covariates with covariates coefficients functions $\rho_{lps}(r), \forall p = 1, ..., P, s = 1, ..., S$, and (v) the unobserved heterogeneity at $l^{\text{th}}$ coefficient level due to the influence of unobserved/unknown factors shared within each unit $i$, captured by continuous latent variable, i.e., $\gamma_{li} \sim N(0, \sigma_{\gamma l}^2)$ where $\sigma_{\gamma l}^2$ is the variance of latent variable.

The proposed model formulation is generic and several of the exiting degradation performance models can be treated as the special cases of the proposed model. For instance, by neglecting the mixed-type covariates and their potential interaction, i.e., $\beta_{li} = 0$, $\alpha_{ls}(\cdot) = 0$, $\rho_{lps}(\cdot) = 0, \forall i = 1, \dots, N, l = 0, \dots, L, p = 1, \dots, P, s = 1, \dots, S$, the proposed model is reduced into the degradation path model in [17]. For another example, by neglecting the functional covariates and the interaction term as well as latent heterogeneity, i.e., $\alpha_{ls}(\cdot) = 0, \rho_{lps}(\cdot) = 0, \gamma_{li} = 0, \forall i = 1, \dots, N, l = 0, \dots, L, p = 1, \dots, P, s = 1, \dots, S$, the pro- posed model becomes the typical accelerated degradation model with constant stress factor. Moreover, by neglecting scalar covariates, potential interaction and latent heterogeneity, i.e., $\beta_{li} = 0$, $\rho_{lps}(\cdot) = 0, \gamma_{li} = 0, \forall i = 1, \dots, N, l = 0, \dots, L, p = 1, \dots, P, s = 1, \dots, S$, the proposed model becomes the degradation model with functional covariates introduced in [22]. Figure 2 further summarizes the hierarchical structure of the proposed bi-level model. As shown in the graphical representation, the square nodes refer to the observed data, including response data and observed environmental conditions as well as measured material characteristics. Circle nodes represent either the unknown model parameters to be estimated or random variables. Triangle nodes represent deterministic functions. The solid and dashed lines indicate the deterministic and stochastic relationship between the connected nodes, respectively.

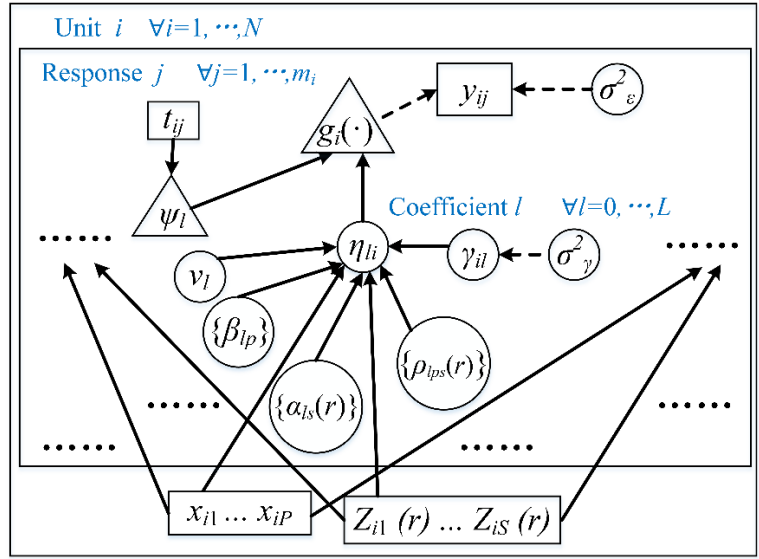

Figure 2: Structure of proposed bi-level degradation performance modeling framework

## 2.2. Material Statistical Descriptor

Among mixed-type covariates in the above formulation, the functional covariates $Z_{is}(r), \forall i = 1, \dots, N, s = 1, \dots, S$ represent critical characteristics of material microstructure, such as spatial heterogeneity, which is known to have indispensable impacts on degradation performance [23]. To evaluate the spatial heterogeneity of material microstructure at microscopic level, advanced sensing technologies, such as transmission electron microscopy (TEM), are often available for the finer scale characterization of material microstructure. Figure 3 (a) gives an example of a TEM image of a tested unit in the accelerated wear test considered in the paper. The gray part and black part represent two different phases of material unit, which have different compositions. As reflected in TEM images with two different colors, such material unit is two-phase and often exhibits spatial heterogeneity and non-uniformity. As compared to the two-phase unit, there is few color contrast or even single color in TEM image of test unit of single phase. The unit of single phase often exhibits homogeneous spatial patterns and uniformity. Figure 3 (a) also illustrates a zoom-in view of spatial patterns of the red dot. To further extract and quantify spatial heterogeneity (or uniformity) patterns of TEM images at reduced complexity, the functional microstructure descriptors [24], such as radial distribution function (RDF) and two-point correlation (TPC) function, are popular choices of correlation-based statistical measures. Unlike scalar covariates which summarize the aggregate information of spatial heterogeneity of material microstructure, RDF and TPC are functional covariates and are capable of capturing spatial heterogeneity patterns more comprehensively. We will elaborate the details of these functional microstructure descriptors as follows.

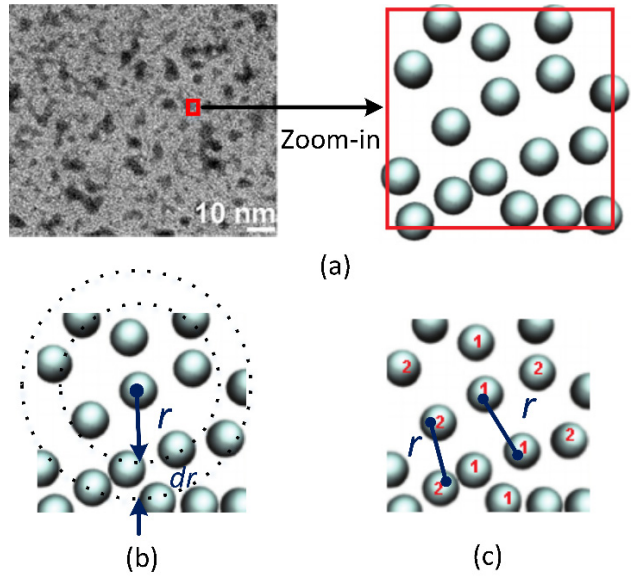

Figure 3: TEM image and its functional descriptor, (a) raw TEM image and zoom-in view, (b) diagram of calculating radial distribution function, (c) diagram of calculating two-point correlation function

Radial distribution function (RDF) is a useful statistical measure to describe how particle density varies as a function of distance from a reference particle [25]. Considering a material test unit of $M$ particles in a volume $V$, RDF can be calculated as $g(\boldsymbol{r}) = \frac{1}{\kappa}\langle\sum_{m\neq 0}\delta(\boldsymbol{r}-\boldsymbol{r}_m)\rangle$ where $\kappa = \frac{M}{V}$ is the average number density of particles and $\langle\cdot\rangle$ is the ensemble averaging operator. $\boldsymbol{r}_m, \forall m = 1, \ldots, M$ refers to particle coordinates and $\delta(\cdot)$ is the Dirac delta function. Particularly, for a test unit with equivalent particles $1, \ldots, M-1$, RDF calculation can be further simplified as $g(\boldsymbol{r}) = V\frac{M-1}{M}\langle\delta(\boldsymbol{r}-\boldsymbol{r}_1)\rangle$. As shown in Figure 3(b), RDF is a distance dependent measure and determines how many particles are within a distance of $r$ and $r + dr$ away from the reference particle. If more particles in a material unit are uniformly distributed, the number of particles within a specified distance from the reference particle over the support range will be similar. This can further be reflected by the RDF values where less sharp changes are involved in the

RDF curve. On the other side, if the particles concentrate on certain area of a test unit, the RDF values of different radius over the support range will become significantly different.

In addition to the RDF measure, we also introduce another measure called two-point correlation function (TPC) [26] to describe microstructure characteristics of material test units. TPC is a statistical measure typically for material test units of two-phase. Considering a two-phase test unit with different material compositions in different phases, we define an indicator function $I^{(h)}(x), h = 1,2$ as $I^{(h)}(x) = \begin{cases} 0, & x \in V_h \\ 1, & x \in \bar{V}_h \end{cases}$ where $V_h$ and $\bar{V}_h$ refer to the region occupied by phase $h$ and the other phase, respectively. TPC then represents the probability of two randomly chosen points $q_1$ and $q_2$ are both in phase $h$, i.e., $S_2^{(h)}(q_1, q_2) = \langle I^{(h)}(q_1) I^{(h)}(q_2) \rangle$ where $\langle \cdot \rangle$ is the ensemble averaging operator over the support range of a test unit. Particularly, when a material test unit is statistically homogeneous, TPC can be calculated as $S_2^{(h)}(q_1, q_2) = \langle I^{(h)}(q_1) I^{(h)}(q_1 + r) \rangle$ where $r$ is specified spatial distance, as depicted in Figure 3 (c). When the particles of same phase are uniformly distributed over a test unit, TPC values of different distances over the support range tend to be similar.

The aforementioned correlation method based material descriptors, such as RDF and TPC, are distance dependent functions and can be used to describe the spatial heterogeneity patterns of material microstructure effectively. The curve shape of these descriptors can be used to differentiate the non-uniformity structure such as clustering pattern from the uniform material structure. With advanced sensing technologies, these functional descriptors become available and can help to study the impacts of material characteristics on the degradation process.

## 2.3. Model Estimation

Considering a population of $N$ deteriorating units are tested and $m_i$ degradation performance measurements are collected on each unit $i$, $i=1,...,N$, the available data can be expressed as $\bm{D} = \{y_{ij}, t_{ij}, \bm{x}_i, Z_{is}(\cdot), \forall i = 1, ..., N, j = 1, ..., m_i, s = 1, ..., S\}$. Let $\bm{\gamma}_i$ be a vector of latent variables of length $L+1$ with $\bm{\gamma}_i \sim N(0, \Sigma_\gamma)$ where $\Sigma_\gamma = [\sigma_{\gamma 0}^2, ..., \sigma_{\gamma L}^2]\mathbf{I}$. We further denote a set of unknown model parameters as $\bm{\Theta} = \{v_l, \bm{\beta}_l, \alpha_{ls}(\cdot), \rho_{lps}(\cdot), \sigma_\varepsilon^2, \forall l = 0, ..., L, p = 1, ..., P, s = 1, ..., S\}$. Suppose the support space of extracted functional covariates from previous section is defined as $\mathbb{R}_s = [0, R]$, the original marginal likelihood function then becomes

$$L(\Theta, \{\gamma_i\}_{i=1}^N \mid D) = \prod_{i=1}^N \int \prod_{j=1}^{m_i} p(y_{ij} \mid \Theta, \gamma_i) p(\gamma_i) d\gamma_i$$

$$\propto \prod_{i=1}^N \int \cdots \int |\sigma_\varepsilon^2 \mathbf{I}_{m_i}|^{-\frac{1}{2}} \exp\{-\frac{1}{2\sigma_\varepsilon^2} \sum_{j=1}^{m_i} (y_{ij} - \sum_{l=0}^L [v_l + \bm{\beta}_l^T \bm{x}_i + \sum_{s=1}^S \int_0^R \alpha_{ls}(r) Z_{is}(r) dr$$

$$+ \sum_{p=1}^P x_{ip} \left( \sum_{s=1}^S \int_0^R \rho_{lps}(r) Z_{is}(r) dr \right) + \gamma_{li}] \phi_l(t_{ij}))^2\} d\gamma_{0i} \ldots d\gamma_{Li} \qquad (2)$$

where $|\cdot|$ refers to the matrix determinant operator and $\mathbf{I}_{m_i}$ is $m_i \times m_i$ identity matrix. As shown in the above likelihood function, the intrinsic infinite dimensionality of functional data [27] makes the parameters estimation mathematically intractable. Besides, as shown in Eq. (2), the latent factors will be integrated out via the marginal approach and cannot be estimated. To estimate the model parameters and latent factors, we need to address these two key challenges. The solution details are elaborated as follows.

The functional data can be treated as a realization of stochastic process and intrinsically involves infinite dimensionality issue. To address such issue, we employ approximation method to reduce the dimensionality of functional data and facilitate parameter estimation. Based on Mercer's theorem, the covariance matrix of functional data can be expressed by orthogonal eigenfunctions and ordered nonnegative eigenvalues [27]. With these eigenfunctions, we can apply Karhunen-Loeve expansion [28] on the centered functional covariates as well as the coefficient functions, and express them as linear combinations of the complete orthogonal basis functions, i.e., $Z_{is}(r) = \sum_{k=1}^{\infty} c_{isk}\psi_k(r), \alpha_{ls}(r) = \sum_{k=1}^{\infty} b_{lsk}\psi_k(r)$ and $\rho_{lps}(r) = \sum_{k=1}^{\infty} b'_{lpsk}\psi_k(r), \forall i = 1, \dots, N, l = 0, \dots, L, p = 1, \dots, P, s = 1, \dots, S$ where $c_{isk}, b_{lsk}$ and $b'_{lpsk}$ are known as functional principle component scores of functional data [27]. Since the eigenvalues of covariance operator of functional data decrease and finally approximate to 0, it is often sufficient to use a small number of eigenfunctions whose eigenvalues are significantly nonzero to accurately approximate the functional data. The number of finite basis functions can be determined efficiently by the fraction of variance explained (FVE) in practice [29]. With the truncated $K$ basis functions, the centered functional covariates and the coefficient functions can then be approximated by $Z_{is}(r) \approx \sum_{k=1}^{K} c_{isk}\psi_k(r), \alpha_{ls}(r) \approx \sum_{k=1}^{K} b_{lsk}\psi_k(r)$ and $\rho_{lps}(r) \approx \sum_{k=1}^{K} b'_{lpsk}\psi_k(r)$. The model parameters then becomes $\boldsymbol{\Theta} = \{v_l, \boldsymbol{\beta}_l, b_{lsk}, b'_{lpsk}, \sigma_\varepsilon^2, \forall l = 0, \dots, L, p = 1, \dots, P, s = 1, \dots, S, k = 1, \dots, K\}$. Consequently, the joint likelihood can be rewritten as

$$L(\boldsymbol{\Theta}, \{\boldsymbol{\gamma}_i\}_{i=1}^N \mid D) \propto \prod_{i=1}^{N} \int \cdots \int |\sigma_\varepsilon^2 \mathbf{I}_{m_i}|^{-\frac{1}{2}} \exp\{-\frac{1}{2\sigma_\varepsilon^2} \sum_{j=1}^{m_i} (y_{ij} - \sum_{l=0}^{L} [v_l + \boldsymbol{\beta}_l^T \boldsymbol{x}_i + R(\sum_{s=1}^{S}\sum_{k=1}^{K} b_{lsk}c_{isk}) + R(\sum_{p=1}^{P} x_{ip}(\sum_{s=1}^{S}\sum_{k=1}^{K} b'_{lpsk}c_{isk})) + \gamma_{li}]\phi_l(t_{ij}))^2\} d\gamma_{0i} \dots d\gamma_{Li} \quad (3)$$

In the above likelihood function, the infinite dimensionality issue of functional data is resolved and it is tractable to estimate $\boldsymbol{\Theta}$. We denote $\boldsymbol{\Lambda}_i$ as $m_i \times (L+1)$ design matrix of latent heterogeneity where matrix details of $\boldsymbol{\Lambda}_i$ are described in appendix A. We then denote $\boldsymbol{\Omega}_i$ as $m_i \times U$ design matrix of observed heterogeneity and denote $\boldsymbol{\zeta}$ as the corresponding coefficient vector of length $U$ where $U = (L+1)(1+P+SK+PSK)$. $\boldsymbol{\Omega}_i$ can be manifested as $\boldsymbol{\Omega}_i = (\boldsymbol{\Lambda}_i \quad \mathbf{A}_{2i} \quad \mathbf{A}_{3i} \quad \mathbf{A}_{4i})$ where matrix details of $\mathbf{A}_{2i}, \mathbf{A}_{3i}$ and $\mathbf{A}_{4i}$ are presented in appendix A. The coefficient vector of observed heterogeneity is written as $\boldsymbol{\zeta} = [\boldsymbol{v}^T, \boldsymbol{\beta}_0^T, \dots, \boldsymbol{\beta}_L^T, \boldsymbol{b}_{01}^T, \dots, \boldsymbol{b}_{LS}^T, \boldsymbol{b}'^T_{011}, \dots, \boldsymbol{b}'^T_{LPS}]^T$ where vector details are described in appendix A. The vector form of the proposed model can then be expressed as $\boldsymbol{y}_i = \boldsymbol{\Omega}_i \boldsymbol{\zeta} + \boldsymbol{\Lambda}_i \boldsymbol{\gamma}_i + \boldsymbol{\varepsilon}_i, i = 1, \dots, N$. The compact likelihood function is given by

$$L(\boldsymbol{\Theta}, \{\boldsymbol{\gamma}_i\}_{i=1}^N \mid D) \propto \prod_{i=1}^{N} \int \cdots \int |\sigma_\varepsilon^2 \mathbf{I}_{m_i}|^{-\frac{1}{2}} \exp(-\frac{1}{2\sigma_\varepsilon^2} \|\boldsymbol{y}_i - \boldsymbol{\Omega}_i \boldsymbol{\zeta} - \boldsymbol{\Lambda}_i \boldsymbol{\gamma}_i\|^2) d\gamma_{0i} \dots d\gamma_{Li}$$
$$(4)$$

In the above marginal approach, the latent variables $\boldsymbol{\gamma}_i, i = 1, \dots, N$ will be integrated out and cannot be estimated. To address the estimation issue of latent factors, We employ data augmentation technique [30] and introduce the complete data, i.e., $\boldsymbol{D}^* = \{\boldsymbol{D}, \{\boldsymbol{\gamma}_i\}_{i=1}^N\}$. The model parameters can be specified as $\boldsymbol{\Theta} = \{\boldsymbol{\zeta}, \sigma_\varepsilon^2, \Sigma_\gamma\}$. Based on the augmented data $\boldsymbol{D}^*$, the joint likelihood can be derived as

$$L(\Theta \mid D^*) = \prod_{i=1}^{N} p(\mathbf{y}_i \mid \Theta, \gamma_i) p(\gamma_i \mid \Theta)$$

$$\propto \prod_{i=1}^{N} \left( |\sigma_\varepsilon^2 \mathbf{I}_{m_i}|^{-\frac{1}{2}} |\Sigma_\gamma|^{-\frac{1}{2}} \exp(-\frac{1}{2}\gamma_i^T \Sigma_\gamma^{-1} \gamma_i) \right) \cdot \exp\left(-\frac{1}{2\sigma_\varepsilon^2} \|\mathbf{y} - \Omega\zeta - \Lambda\gamma\|^2 \right) \quad (5)$$

where $\|\cdot\|$ is Euclidean norm operator. $\mathbf{y} = [\mathbf{y}_1^T, \ldots, \mathbf{y}_N^T]^T$ is a vector of length $\sum_{i=1}^{N} m_i$ representing the degradation performance outputs for all units. $\Omega = (\Omega_1^T \quad \ldots \quad \Omega_N^T)^T$ is a matrix of dimension $\sum_{i=1}^{N} m_i$ by $U$ representing the design matrix of observed heterogeneity for all units. $\Lambda = \begin{pmatrix} \Lambda_1 & 0 & \ldots & 0 \\ 0 & \Lambda_2 & \ldots & 0 \\ 0 & 0 & \ldots & \Lambda_N \end{pmatrix}$ is a matrix of dimension $\sum_{i=1}^{N} m_i \times (L+1)N$ representing the design matrix of latent heterogeneity for all units. $\gamma = [\gamma_1^T, \ldots, \gamma_N^T]^T$ is a vector of length $(L+1)N$ representing the latent variables among all units. Further, the log likelihood of augmented data can be written as $l(\Theta|D^*) = l_1(\zeta, \sigma_\varepsilon^2|D^*) + l_2(\Sigma_\gamma|D^*)$. $l_1(\zeta, \sigma_\varepsilon^2|D^*)$ can be manifested as $l_1(\zeta, \sigma_\varepsilon^2|D^*) = -\frac{\sum_{i=1}^{N} m_i}{2}\log(\sigma_\varepsilon^2) - \frac{1}{2\sigma_\varepsilon^2}\left(\|\mathbf{y} - \Omega\zeta\|^2 + \sum_{i=1}^{N} Tr(\Lambda_i^T \Lambda_i \gamma_i \gamma_i^T) - 2(\mathbf{y} - \Omega\zeta)^T \Lambda\gamma\right)$ where $Tr(\cdot)$ is the trace operator. $l_2(\Sigma_\gamma|D^*)$ can be expressed as $l_2(\Sigma_\gamma|D^*) = -\frac{N}{2}\log|\Sigma_\gamma| - \frac{1}{2}\sum_{i=1}^{N} \gamma_i^T \Sigma_\gamma^{-1} \gamma_i$. Given that $\gamma$ is known, $\zeta$ and $\sigma_\varepsilon^2$ can be obtained by maximizing $l_1(\zeta, \sigma_\varepsilon^2|D^*)$ and $\Sigma_\gamma$ can be estimated by maximizing $l_2(\Sigma_\gamma|D^*)$. However, the latent factors are unknown in real world problem. Thus, we employ EM technique [30, 31] to develop the estimation algorithm to jointly estimate unknown model parameters and latent factors. At iteration $\tau$, the Expectation step yields the conditional expectation of $l(\Theta|D^*)$, i.e., $Q(\Theta, \Theta^{(\tau-1)}) = \mathbb{E}_{S(\gamma)|D,\Theta^{(\tau-1)}}[l(\Theta|D^*)]$ where $S(\gamma) = (\gamma_1, \ldots, \gamma_N, \gamma_1 \gamma_1^T, \ldots, \gamma_N \gamma_N^T)$ is a set of individual statistics and $\Theta^{(\tau-1)}$ is a collection of all obtained model parameters at iteration $\tau - 1$. The Q-function can be explicitly written as

$$Q(\Theta, \Theta^{(\tau-1)}) \propto -\frac{\sum_{i=1}^{N} m_i}{2} \log(\sigma_\varepsilon^2) - \frac{1}{2\sigma_\varepsilon^2}(\|\mathbf{y} - \Omega\zeta\|^2$$
$$+ \sum_{i=1}^{N} Tr(\Lambda_i^T \Lambda_i \mathbb{E}[\gamma_i \gamma_i^T \mid D, \Theta^{(\tau-1)}]) - 2(\mathbf{y} - \Omega\zeta)^T \Lambda \mathbb{E}[\gamma \mid D, \Theta^{(\tau-1)}])$$
$$- \frac{N}{2}\log|\Sigma_\gamma| - \frac{1}{2}\sum_{i=1}^{N} \mathbb{E}[\gamma_i^T \Sigma_\gamma^{-1} \gamma_i \mid D, \Theta^{(\tau-1)}] \quad (6)$$

where $Tr(\cdot)$ is the trace operator. The corresponding conditional expectation $\mathbb{E}[\gamma_i|D, \Theta^{(\tau-1)}]$ and $\mathbb{E}[\gamma_i \gamma_i^T|D, \Theta^{(\tau-1)}]$ can be explicitly obtained as

$$\mathbb{E}[\gamma_i \mid D, \Theta^{(\tau-1)}] = \mu_i^{(\tau-1)} = \frac{1}{\sigma_\varepsilon^{2(\tau-1)}} V_i^{(\tau-1)} \Lambda_i^T (\mathbf{y}_i - \Omega_i \zeta^{(\tau-1)}), \forall i = 1, \ldots, N$$
$$\mathbb{E}[\gamma_i \gamma_i^T \mid D, \Theta^{(\tau-1)}] = V_i^{(\tau-1)} + \mu_i^{(\tau-1)} \mu_i^{(\tau-1)T}, i = 1, \ldots, N \quad (7)$$

where $V_i^{(\tau-1)} = \left(\Sigma_\gamma^{-1(\tau-1)} + \frac{1}{\sigma_\varepsilon^{2(\tau-1)}} \Lambda_i^T \Lambda_i\right)^{-1}$. The derivation details of conditional expectation are presented in appendix B.1. We can further simplify the conditional expectation as $\mathbb{E}[\gamma|D, \Theta^{(\tau-1)}] = \mu^{(\tau-1)} = [\mu_1^{(\tau-1)T}, \ldots, \mu_N^{(\tau-1)T}]^T$. With the calculated Q-function at current iteration, the Maximization step achieves the maximization of $Q(\Theta, \Theta^{(\tau-1)})$. The model parameters at iteration $\tau$ can then be updated by

$\boldsymbol{\Theta}^{(\tau)} = \arg\max_{\boldsymbol{\Theta}} Q(\boldsymbol{\Theta}, \boldsymbol{\Theta}^{(\tau-1)})$. It is mathematically tractable to maximize $Q(\boldsymbol{\Theta}, \boldsymbol{\Theta}^{(\tau-1)})$ and the closed form of estimated model parameters at iteration $\tau$ can be expressed as

$$\hat{\zeta}^{(\tau)} = (\Omega^T \Omega)^{-1} \Omega^T \left( y - \Lambda \mu^{(\tau-1)} \right) \quad (8)$$

$$\hat{\Sigma}_\gamma^{(\tau)} = \frac{1}{N} \sum_{i=1}^{N} \left( (\hat{\Sigma}_\gamma^{(\tau-1)})^{-1} + \frac{1}{\hat{\sigma}_\epsilon^{2(\tau-1)}} \Lambda_i^T \Lambda_i \right)^{-1} + \mu_i^{(\tau-1)} \mu_i^{(\tau-1)T} \right) \quad (9)$$

$$\hat{\sigma}_\epsilon^{2(\tau)} = \frac{1}{\sum_{i=1}^{N} m_i} \left( \| y - \Omega \hat{\zeta}^{(\tau)} \|^2 - 2 \sum_{i=1}^{N} (y_i - \Omega_i \hat{\zeta}^{(\tau)})^T \Lambda_i \mu_i^{(\tau-1)} \right.$$
$$\left. + \sum_{i=1}^{N} \mathrm{Tr}(\Lambda_i^T \Lambda_i ((\hat{\Sigma}_\gamma^{(\tau)})^{-1} + \frac{1}{\hat{\sigma}_\epsilon^{2(\tau-1)}} \Lambda_i^T \Lambda_i)^{-1} + \mu_i^{(\tau-1)} \mu_i^{(\tau-1)T})) \right) \quad (10)$$

The derivation details of the above estimation procedure are provided in appendix B.2. The estimation algorithm of the proposed model is summarized in Algorithm 1. In each iteration, the conditional expectations are computed first and the parameters are then updated sequentially via maximizing Q-function. The algorithm updates parameters iteratively till the maximum iteration $\tau_{\max}$ is achieved. The EM estimation technique greatly simplifies the computational difficulties in joint likelihood maximization. The merits of data augmentation method enables the closed-form parameter updating procedure. It is proved that the parameter estimates will converge to maximum likelihood estimates when the updating iteration increases [31].

---
**Algorithm 1** Parameters estimation procedure of the proposed model

**Initialization:** $\Theta^{(0)} = \{\zeta^{(0)}, \sigma_\epsilon^{2(0)}, \Sigma_\gamma^{(0)}\}$

**procedure** UPDATEESTIM
    **for** $\iota \leftarrow 1, ..., \tau_{\max}$ **do**
        Compute $E[\gamma_i \mid D, \Theta^{(\iota-1)}]$ and $E[\gamma_i \gamma_i^T \mid D, \Theta^{(\iota-1)}]$ based on Eq. (7)
        Derive parameter estimates sequentially by maximizing $Q(\Theta, \Theta^{(\iota-1)})$
            1. Obtain $\hat{\zeta}^{(\iota)}$ based on Eq. (8)
            2. Obtain $\hat{\Sigma}_\gamma^{(\iota)}$ based on Eq. (9)
            3. Obtain $\hat{\sigma}_\epsilon^{2(\iota)}$ based on Eq. (10)
    **end for**
**end procedure**

---

## 3. Real Case Study

### 3.1. Experimental Data Description

To illustrate the proposed modeling framework and to further evaluate its prediction performance as well as model interpretation capability, we provide a real case study to analyze the tribological degradation performance data of copper alloys. Due to their exceptional mechanical properties and high strength, copper alloys have been widely considered in various mission and safety critical components (e.g., aircraft bearings and bushings) and systems (e.g., drilling and mining systems). The tribological degradation of copper alloys may be influenced by both external factors, such as load conditions, and internal factors, such as material microstructure information. We will utilize the proposed degradation performance model to quantify the influences of these observed factors as well as unobserved heterogeneity within each individual test unit.

Specifically, accelerated wear tests of Cu-Ni-Sn alloys are carried out at elevated load conditions using the Koehler K93500 pin-on-disc tester [23], as shown in Figure 4. The test units consist of both as-received and annealed material specimens of Cu-Ni-Sn alloys. As compared to the as-received test units, the microstructure and physical/chemical properties of annealed test units are often altered considerably through the annealing process, and therefore the corresponding tribological degradation performance may also differ. For each test unit, its tribological degradation performance outputs (i.e., height loss in um) are measured over time (in seconds) by a linear variable displacement transducer (LVDT). Figure 5 shows degradation performance observations of four test units with two different material types under different load conditions. The graphical visualization implies that both the internal factors (e.g., material types) and external factors (e.g., load conditions) play important roles in the degradation performance outputs and there is a need to explicitly quantify their effects.

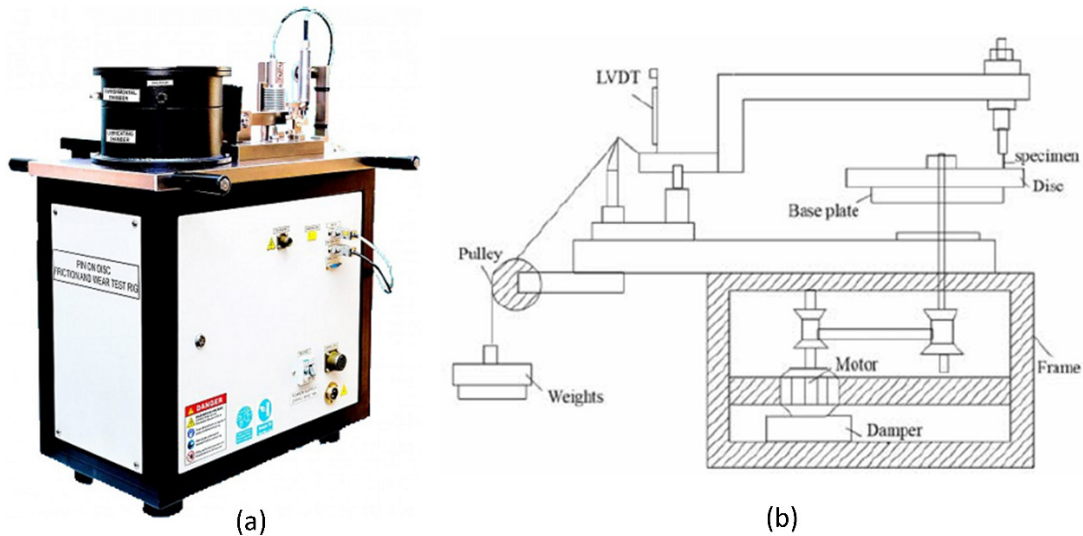

Figure 4: The testing environment of the investigated accelerated wear test, (a) the testing equipment, (b) the diagram of wear test

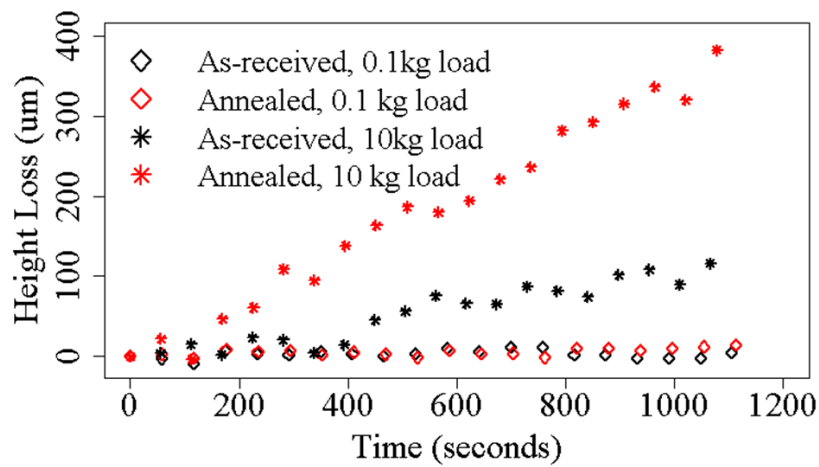

Figure 5: Degradation performance outputs of four test units under different material types (e.g., as-received and annealed) and load conditions (e.g., 0.1 kg and 10 kg)

With the advancement of material sensing and characterization techniques, it becomes readily available nowadays to obtain and extract more detailed information for test units with different material types. In this paper, we utilize transmission electron microscopy (TEM) technique to characterize the microstructure of both as-received and annealed test units. The TEM images embrace useful information about material microstructure at finer scale. Particularly, we use grayscale TEM images at a nanometer-or-less length scale (20 nm) to characterize the microstructure properties of different types of copper alloys. Figure 6 shows two TEM images of as-received and annealed test units, respectively, which have significant visual difference. The texture pattern of the annealed test unit in Figure 6 (b) has less spatial heterogeneity than that of the as-received test unit, indicating a more homogeneous microstructure with less distinct crystal structure and chemical composition. Such microstructure difference among different types of copper alloys is essentially due to the influence of the annealing process. After annealing at higher temperature, the primary two-phase microstructure in the as-received test unit has been converted into a single phase with alerted mechanical properties.

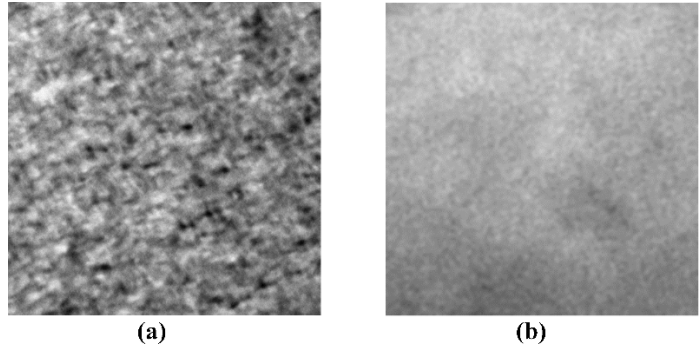

(a)           (b)

Figure 6: TEM micrographs of two different material types: (a) as-received copper alloy, (b) annealed copper alloy

*3.2. Functional Covariates Extraction*

As shown in Figure 6, the microstructures of the test units with different material types exhibit different spatial heterogeneity patterns. To capture the spatial heterogeneity with rich spatial information, two functional microstructural descriptors, namely TPC and RDF, are considered. As described in Section. 2.2, the functional covariate $Z(r)$ is used to represent the material microstructure information where $r$ is the distance measured at spatial scale in pixel. The distance measure for TPC function ranges from 0 to 200 pixels and the distance measure for RDF is between 0 and 250 pixels. Figure 7 shows the corresponding functional covariates extracted based on TPC and RDF descriptors. Both descriptors are able to capture and differentiate the spatial heterogeneity patterns between annealed and as-received test units. The TPC (or RDF) value of an annealed test unit is uniformly larger than that of an as-received one at various spatial scales (measured in pixels), indicating a uniformly more homogeneous material microstructure. As the spatial distance r increases, the microstructures of both test units tend to become less homogeneous (with a smaller TPC or RDF value). When the spatial distance $r$ is larger than 25 pixels, the TPC (or RDF) value of as-received test unit tends to approach 0, indicating that the cluster size in its TEM image is typically no larger than 25 pixels. Further, TPC is more sensitive than RDF in capturing and differentiating spatial heterogeneity of annealed unit at finer scale (e.g., smaller than 15 pixels).

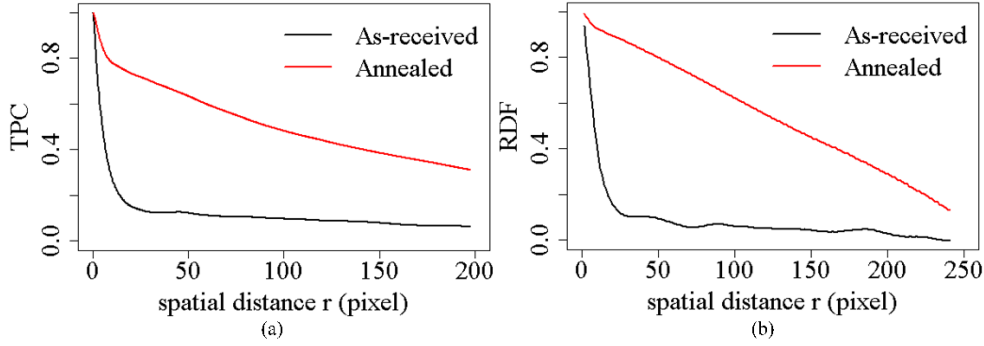

Figure 7: Functional covariates extracted based on different functional microstructural descriptors: (a) TPC, (b) RDF

*3.3. Performance Comparison With Alternative Modeling Approaches*

We first apply centering to the raw degradation performance data such that their baseline degradation performance (at $t = 0$) has been centered around 0, i.e., $\eta_{0i} = 0, \forall i = 1, \ldots, 12$. As shown in Figure 5, the degradation performance outputs exhibit approximately linear degradation path and thus we consider the first order polynomial basis function, i.e., $\phi_1(t_{ij}) = t_{ij}, \forall i = 1, \ldots, 12, j = 1, \ldots, 20$. We then decompose the basis coefficient $\eta_{1i}$ into several parts, as described in Section. 2.1. We use the extracted functional covariate $Z_{i1}(r)$ for each test unit $i, \forall i = 1, \ldots, 12$, as described in previous section, to represent the unit-specific microstructure pattern. In addition, we consider the scalar covariate $x_{i1}$ to represent the load condition exerted on test unit $i$. By incorporating both $Z_{i1}(r)$ and $x_{i1}$ as well as their interaction term into the proposed model, we further introduce a unit-specific latent variable $\gamma_{1i}$ to capture the unobserved heterogeneity within each test unit $i$. The model specification using higher-order polynomial basis functions is also investigated and compared, which will be shown later in this section. Given a sample of test units available with observed degradation performance data (e.g., the height loss in um unit), we estimate the proposed model based on the estimation procedure described in Section. 2.3. For basis function approximation in handling the infinite dimensionality of functional covariate during estimation, we select the optimal number of orthogonal basis functions based on the criterion of percentage of functional variation explained (FVE). The final number of components for TPC functional approximation is 2 and the basis order for RDF functional approximation is 3.

With the selected number of basis functions and the estimated models, we further evaluate the models with different functional descriptors (e.g., TPC and RDF) and compare their prediction performances. We partition the data into training (first 80% of degradation observations of each test unit) and testing data sets. We employ 5-fold cross-validation (CV) to evaluate and compare the prediction performance of the two models. The total CV error of all test units using model with TPC descriptor is smaller than the total CV error of RDF-based model. Thus, we select TPC as microstructural descriptor to evaluate the performance of proposed modeling framework and further to compare it with other alternative models.

To demonstrate the superior modeling performance of the proposed modeling framework which incorporates mixed-type (i.e., both scalar and functional) covariates, their interaction as well as unobserved heterogeneity, we investigate and compare several alternative modeling approaches at various level of model complexity. First, to investigate the importance of mixed-type covariates and their interaction, we consider the following set of alternative modeling approaches, namely (i) Model 1 which only considers scalar covariate of load condition, (ii) Model 2 which only considers functional covariate of material microstructure, (iii) Model 3 which considers mixed-type covariates but fails to consider their interaction, and (iv) Model 4 which considers both mixed-type covariates and their interaction but fails to consider unobserved heterogeneity. Among the first four modeling approaches, Model 4 is the most complex one. Based on Model 4, we further adjust its modeling complexity from the following three aspects, namely (i) adding higher order basis function in Model 5; (ii) replacing functional covariate with scalar covariate to represent material microstructure in Model 6; and (iii) incorporating latent variable to represent unobserved heterogeneity in Model 7 (i.e., the proposed model). It is noticed that, for Model 6, we consider the scalar covariate of fractal dimension $D\alpha$ to capture material microstructure, which is a popular choice of feature extraction method and often considered in microstructure image analysis [32]. Table 1 summarizes the differences between the proposed approach and 6 alternative modeling approaches.

Table 1: Different degradation performance modeling approaches

| Model | Observed factors | | | | Unobserved heterogeneity | Order of polynomial basis | |
|---|---|---|---|---|---|---|---|
| | Material microstructure | | Load condition | Interaction | | first order | first and second orders |
| | Functional | Scalar | Scalar | | | | |
| Model 1 | | | ✓ | | | ✓ | |
| Model 2 | ✓ | | | | | ✓ | |
| Model 3 | ✓ | | ✓ | | | ✓ | |
| Model 4 | ✓ | | ✓ | ✓ | | ✓ | |
| Model 5 | ✓ | | ✓ | ✓ | | ✓ | ✓ |
| Model 6 | | ✓ | ✓ | ✓ | | ✓ | |
| Model 7 (Proposed) | ✓ | | ✓ | ✓ | ✓ | ✓ | |

To comprehensively evaluate and compare the modeling performance of the above models, we consider the following different evaluation criteria, namely (i) model goodness-of-fit criteria, e.g., R-squared and data log-likelihood; (ii) model selection criteria, e.g., Akaike Information Criterion (AIC) and Bayesian Information Criterion (BIC), and (iii) model prediction criteria, e.g., mean squared error (MSE) of both training and test data sets. The model selection criteria of AIC and BIC can be unified as $C(\Theta) = -2 \ln L(D|\widehat{\Theta}, \{\hat{\gamma}_i\}_{i=1}^N) + pk$, where $L(D|\widehat{\Theta}, \{\hat{\gamma}_i\}_{i=1}^N)$ is the maximized joint likelihood and $pk$ is the penalty term to leverage the model complexity. In particular, $p$ is the total number of estimated parameters in the model, $k=2$ (for AIC) and $k=\ln N$ (for BIC).

Based on the above criteria, Table 2 summarizes the results of performance comparison. Among all models, the proposed model achieves the best performance, in terms of both the largest goodness-fit, the smallest AIC/BIC and the smallest MSE values. Several additional implications and discussions can be obtained as follows. First, by comparing Model 3 with Models 1 and 2, the model which only considers either external factor of environmental conditions (e.g., Model 1) or internal factor of material types (e.g., Model 2) has worse goodness-of-fit and prediction performance than the model which considers both internal and external factors (e.g., Model 3). Further, there exists interaction between these two factors and capturing such interaction (e.g., Model 4) will further improve the modeling performance. This implies the importance of incorporating both internal and external factors (as well as their potential interaction) during the degradation performance modeling. Second, by comparing Models 4 and 5, we explore the impact of higher order basis function on modeling performance improvement. Based on the training and testing MSE values, model with extra quadratic term (e.g., Model 5) has overfitting issue. Thus, in this study, the complex curvilinear model specification with higher order basis function will not provide additional modeling benefits and need to be avoided. Third, by comparing Models 4 and 6, Model 4 which considers mixed-type covariates (e.g., functional covariate of material microstructure and scalar covariate of load conditions) has superior modeling performance than Model 6 which only considers single-type covariates (e.g., scalar covariates). This emphasizes the benefits of incorporating mixed-type covariates in improving degradation performance modeling accuracy. The functional covariate carries richer material microstructure information than the conventional scalar covariate. Last, by comparing the proposed model which considers unobserved heterogeneity (e.g., Model 7) and Model 4 which fails to consider unobserved heterogeneity, both model goodness-of-fit and prediction results indicate the importance of considering both observed factors' influences and unobserved heterogeneity in developing the degradation performance model. The predicted degradation performance based on different models (e.g., Models 1-7) and the observed degradation performance of a single test unit are also displayed in Figure 8. As compared to the alternative modeling approaches (e.g., Models 1-6), the predicted degradation performance values based on the proposed model (e.g., Model 7) are closer to the actual degradation performance observations over time.

Table 2: Model comparison results

| Model | Results | | | | | Model Structure |
|---|---|---|---|---|---|---|
| | $R^2$ | $\log L$ | AIC | BIC | $(\bar{\Delta}_{\text{train}}, \bar{\Delta}_{\text{test}})$ | |
| Model 1 | 0.7303 | -937.4 | 1880.8 | 1890.6 | (1019.2, 3779.6) | $x$ |
| Model 2 | 0.5672 | -982.3 | 1972.6 | 1985.6 | (1626.8, 6646.2) | $Z(r)$ |
| Model 3 | 0.9269 | -811.1 | 1632.1 | 1648.4 | (273.3, 1067.6) | $x + Z(r)$ |
| Model 4 | 0.9736 | -712.1 | 1438.3 | 1461.1 | (97.5, 248.9) | $x \times Z(r)$ |
| Model 5 | 0.9735 | -712.1 | 1440.1 | 1466.2 | (97.5, 253.6) | $\phi_1(t_{ij}) + \phi_2(t_{ij})$ |
| Model 6 | 0.7304 | -936.3 | 1882.7 | 1898.965 | (1007.9, 3748.8) | $Z$ |
| Model 7 | 0.9681 | -698.7 | 1413.3 | 1439.4 | (72.4, 156.3) | $x \times Z(r) + \gamma$ |

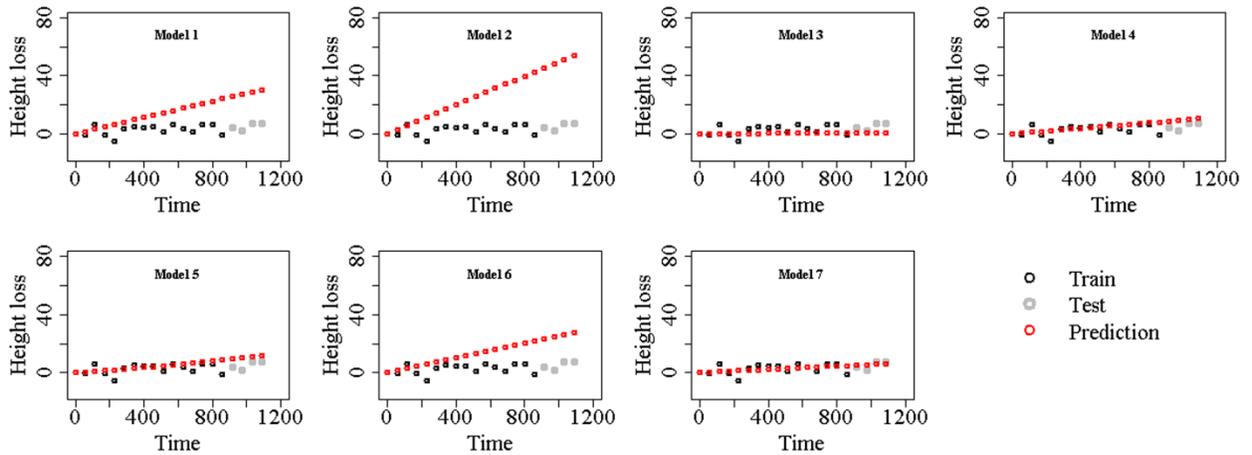

Figure 8: Prediction performance comparison among different models

### 3.4. Model Interpretation

Based on the above performance evaluation and comparison, the superior prediction performance of proposed modeling framework is demonstrated. We further investigate the model interpretability of the proposed work. The effects of different types of covariates on tribological degradation are discussed as below. The point estimates of model coefficients and their p-values are summarized in Table. 3. As shown in Table 3, both $\hat{v}_1$ and $\hat{\beta}_{11}$ are significant at a significant level of 0.05. $\hat{v}_1$ captures the tribological degradation rate of studied alloy in absence of the influences of mixed-type covariates. A positive value of

Table 3: Model parameter estimation results

| Parameter | Point Estimate | P-value | Parameter | Point Estimate | P-value |
|---|---|---|---|---|---|
| $v_1$ | 0.01633 | 0.0038 | $\beta_{11}$ | 0.02046 | <.0001 |
| $b_{111}$ | 0.000025 | 0.0235 | $b'_{1111}$ | 0.00002 | <.0001 |
| $b_{112}$ | -0.00047 | 0.4234 | $b'_{1112}$ | 0.000326 | 0.3298 |

$\hat{v}_1$ indicates an increasing trend of degradation performance (i.e., material height loss) of copper alloys over time. $\hat{\beta}_{11}$ quantifies the marginal effects of scalar covariate, i.e., load condition, on degradation performance output of test units. A positive value of $\hat{\beta}_{11}$ indicates that load condition is an effective stress factor in accelerating the tribological degradation process of copper alloys. Further, we investigate the marginal and interaction effect of functional covariates (i.e., TPC curves) of material microstructures on degradation performance output. As shown in Eq. (1b), $\int_0^R \alpha_{11}(r) Z_{i1}(r)\, dr$ captures the contributing marginal effects of functional covariate $Z_{i1}(r)$ on the slope of degradation performance, which can be written as $R(b_{111}c_{i11} + b_{112}c_{i12})$ via basis function approximation (in Section 2.3). A non-zero value of parameter $b_{111}$ at a significance level of 0.05 indicates that there exists a marginal effect of material microstructures (captured by functional covariate) among different material types. Figure 9 shows the estimated contributing marginal effects, $\int_0^R \alpha_{11}(r) Z_{i1}(r)\, dr$, among different test units with two different types of copper alloys. As shown in Figure 9, the microstructure effects of as-received test units are negative.

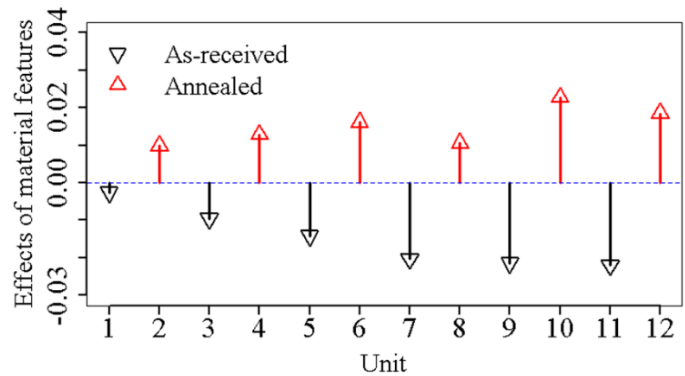

Figure 9: Effects of internal material characteristics among all test units

It indicates that a less homogeneous microstructure yields greater hardness of the materials and ultimately reduces the corresponding tribological degradation rate. On the other side, the microstructure effects of annealed test units are positive, which indicates that a more uniformly homogeneous microstructure tends to accelerate the tribological degradation process of test units. In addition to the significant marginal effect of functional covariate, the non-zero value of parameter $b'_{1111}$ in Table 3 at a significance level of 0.05 also indicates that there exists an interaction effect between material microstructure (captured by functional covariate) and load condition (captured by scalar covariate) on the slope of degradation performance. Specifically, as shown in Eq. (1b), such interaction effect between scalar and functional covariates can be characterized as $x_{i1} \int_0^R \rho_{111}(r) Z_{i1}(r) \, dr$, which can further be approximated as $R x_{i1} (b'_{1111} c_{i11} + b'_{1112} c_{i12})$ (as manifested in Section. 2.3). Figure 10 shows the estimated interaction effect, $x_{i1} \int_0^R \rho_{111}(r) Z_{i1}(r) \, dr$, among different test units. As shown in Figure 10, the interaction effect of material microstructure and load condition for as-received test units are negative. It implies that the degradation performance of as-received test units with less homogeneous microstructure is less sensitive to the accelerated load condition. On the other side, the interaction effect of material microstructure and load condition for annealed test units are positive. It implies that for more homogeneous microstructure, there exists a synergistic effect between material microstructure and accelerated load condition, which can further speed up the accelerated degradation test. All of the above rich model interpretations will inform the reliability engineers at product design and development stage to better identify the most appropriate material types as well as accelerated conditions to improve the efficiency of accelerated testing as well as to improve the reliability performance of test units. In addition to the quantification of the influences of the observed factors, the proposed work is able to quantify the unobserved heterogeneity, captured by $\gamma_{1i}$, within each test unit. As shown in Figure 11, the positive value of unobserved heterogeneity indicates a positive effect on the slope of degradation performance and vice versus. Such unobserved heterogeneity is essentially caused by unobserved/unknown factors shared within each test unit (e.g., the potential effect of accumulated wear debris on specimen contact surface. Such information can inform the reliability engineers to target specific test units which exhibit large unobserved heterogeneity and to investigate the potential contributing factors within each test unit via further data collection and analysis.

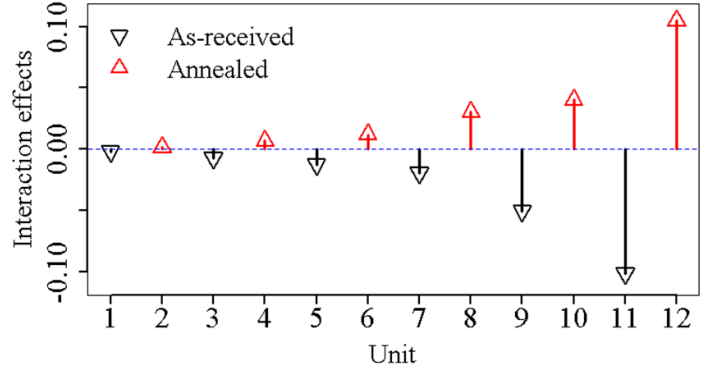

Figure 10: Interaction effect of external load condition and internal material microstructure among all test units

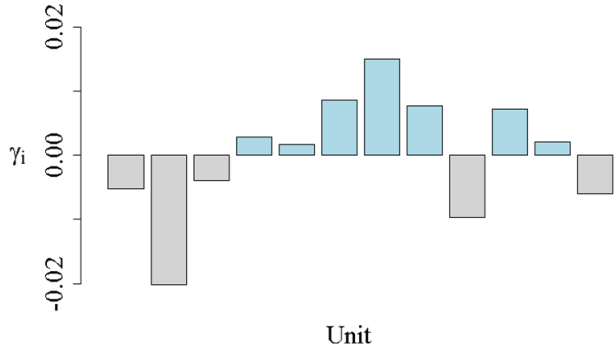

Figure 11: Quantification of individual latent heterogeneity

## 4. Conclusion

In this paper, we propose a degradation performance modeling framework with both generic model formulation and effective estimation algorithm to characterize heterogeneous degradation data with covariates. The proposed model formulation allows the incorporation of both mixed-type covariates and latent heterogeneity for improving the prediction accuracy of degradation performance. The mixed-type covariates include both (i) the scalar covariates, which represent the accelerated/environmental conditions of test units; and (ii) the functional covariates, which represent the material microstructure characteristics of test units. The developed estimation algorithm further allows the joint quantification of influences of mixed-type covariates as well as unobserved heterogeneity. In particular, basis function approximation technique is employed to address the infinite dimensionality issue of model estimation when functional covariate is involved. Data augmentation technique is further employed to simultaneously estimate both the observed effects of mixed-type covariates as well as unobserved heterogeneity within each unit which is captured by the unit specific latent variable. A real case study is also presented to illustrate the proposed modeling framework and evaluate its performance via comprehensively comparing with several alternative degradation performance modeling approaches, such as models only considering scalar or functional covariates, and models neglecting the unobserved heterogeneity. The proposed work demonstrates its superior prediction performance via using different performance metrics, such as goodness-of-fit, model selection criteria and model prediction criteria. With improved prediction accuracy of degradation performance, the proposed model also provides rich model interpretation. Both the marginal effects among mixed-type covariates, their interaction effect and unobserved heterogeneity within each test unit can be jointly quantified explicitly. Such modeling outputs will help to inform the reliability engineers to better identify the most influencing factors for improving reliability performance of product units and design more efficient accelerated tests in response to the varied material characteristics of product units.

## Appendix

*A. Details of vector form of proposed model*

Based on Eq. (3), the degradation performance responses can be fully expressed as

$$y_{ij} = \sum_{l=0}^{L}[\nu_l + \boldsymbol{\beta}_l^T \boldsymbol{x}_i + R\left(\sum_{s=1}^{S}\sum_{k=1}^{K} b_{lsk}c_{isk}\right)$$
$$+ R\left(\sum_{p=1}^{P} x_{ip}\left(\sum_{s=1}^{S}\sum_{k=1}^{K} b'_{lpsk}c_{isk}\right)\right) + \gamma_{li}]\phi_l(t_{ij}) + \epsilon_{ij}, i = 1, \dots, N, j = 1, \dots, m_i$$

where $y_{ij}$ is degradation performance output of test unit $i$ measured at time $t_{ij}$ and $\varepsilon_{ij}$ is the error term. We denote the vector form as $\boldsymbol{y}_i = [y_{i1}, \dots, y_{im_i}]^T$ and $\boldsymbol{\varepsilon}_i = [\varepsilon_{i1}, \dots, \varepsilon_{im_i}]^T$. $\nu_l$ is the population-level average degradation performance at $l^{\text{th}}$ decomposition, $\forall l = 0, \dots, L$. Let $\boldsymbol{\nu} = [\nu_0, \dots, \nu_L]^T$ be a vector of population-level average degradation performance at all decomposition levels. $x_{ip}$ is the $p^{th}$ scalar covariate of unit $i$ and $\boldsymbol{x}_i = [x_{i1}, \ldots, x_{iP}]$ is a vector of all $P$ observed scalar covariates. $\boldsymbol{\beta}_l = [\beta_{l1}, \ldots, \beta_{lP}]^T$ is a vector of the coefficients of total $P$ scalar covariates at $l^{th}$ decomposition, $\forall l = 0, \ldots, L$. $R$ is the range of spatial distance and $c_{isk}$ is the $k^{th}$ basis coefficient for $s^{th}$ functional covariate of unit $i$. $b_{lsk}$ and $b'_{lpsk}$ are $k^{th}$ basis coefficients for coefficient function of $s^{th}$ functional covariate and the interaction terms, respectively. We denote $\boldsymbol{b}_{ls} = [b_{ls1}, \ldots, b_{lsK}]^T$ and $\boldsymbol{b}'_{lps} = [b'_{lps1}, \ldots, b'_{lpsK}]^T, \forall l = 0, \ldots, L, p = 1, \ldots, P, s = 1, \ldots, S$. $\gamma_{li}$ is the latent factor of unit $i$ at $l^{th}$ decomposition and let $\boldsymbol{\gamma}_i = [\gamma_{0i}, \ldots, \gamma_{Li}]^T$ be a vector of latent variables of unit $i$ at all decomposition levels. $\phi_l(\cdot)$ is basis function at $l^{th}$ decomposition. The design matrix of unobserved heterogeneity can be expressed as

$$\Lambda_i = (w_{uv}^{1i}) \in \mathbb{R}^{m_i \times (L+1)}$$

where $w_{uv}^{1i} = \phi_v(t_{iu}), \forall u = 1, \ldots, m_i, v = 0, \ldots, L$ is the element at $u^{th}$ row and $v^{th}$ column. Further, we let $\mathbf{A}_{2i}, \mathbf{A}_{3i}$ and $\mathbf{A}_{4i}$ be the block matrices, i.e.,

$$A_{2i} = (A_{20i} \quad A_{21i} \quad \ldots \quad A_{2Li}) \in \mathbb{R}^{m_i \times (L+1)P}$$

where each block $\mathbf{A}_{2li} = (w_{uv}^{2li}) \in \mathbb{R}^{m_i \times P}, \forall l = 0, \ldots, L$ with the element at $u^{th}$ row and $v^{th}$ column of the submatrix as $w_{uv}^{2li} = x_{iv}\phi_l(t_{iu}), \forall u = u = 1, \ldots, m_i, v = 1, \ldots, P$.

$$A_{3i} = (A_{30i} \quad A_{31i} \quad \ldots \quad A_{3Li}) \in \mathbb{R}^{m_i \times (L+1)SK}$$

where each block $\mathbf{A}_{3li} \in \mathbb{R}^{m_i \times SK}, \forall l = 0, \ldots, L$ can further be expressed as $\mathbf{A}_{3li} = (\mathbf{A}_{3li1} \quad \ldots \quad \mathbf{A}_{3liS})$, $\forall l = 0, \ldots, L$. Each block $\mathbf{A}_{3lis}$ can be written as $\mathbf{A}_{3lis} = (w_{uv}^{3lis}) \in \mathbb{R}^{m_i \times K}, \forall s = 1, \ldots, S$ where $w_{uv}^{3lis} = Rc_{isv}\phi_l(t_{iu}), \forall u = u = 1, \ldots, m_i, v = 1, \ldots, K$ is the element at $u^{th}$ row and $v^{th}$ column of the submatrix.

$$A_{4i} = (A_{40i} \quad A_{41i} \quad \ldots \quad A_{4Li}) \in \mathbb{R}^{m_i \times (L+1)PSK}$$

where $\mathbf{A}_{4li} = (\mathbf{A}_{4li1} \quad \ldots \quad \mathbf{A}_{4liP}) \in \mathbb{R}^{m_i \times PSK}, \forall l = 0, \ldots, L$. Each block $\mathbf{A}_{4lip}, \forall p = 1, \ldots, P$ can be further expressed as $\mathbf{A}_{4lip} = (\mathbf{A}_{4lip1} \quad \ldots \quad \mathbf{A}_{4lipS}) \in \mathbb{R}^{m_i \times SK}$. Finally, the submatrix $\mathbf{A}_{4lips}, \forall s = 1, \ldots, S$ can be written as $\mathbf{A}_{4lips} = (w_{uv}^{4lips}) \in \mathbb{R}^{m_i \times K}$ where $w_{uv}^{4lips} = Rx_{ip}c_{isv}\phi_l(t_{iu}), \forall u = 1, \ldots, m_i, v = 1, \ldots, K$ is the element at $u^{th}$ row and $v^{th}$ column.

With the above block matrices, the design matrix of observed heterogeneity can then be manifested as

$\boldsymbol{\Omega}_i = (\Lambda_i \quad \mathbf{A}_{2i} \quad \mathbf{A}_{3i} \quad \mathbf{A}_{4i}) \in \mathbb{R}^{m_i \times U}$ where $U = (L+1)(1+P+SK+PSK)$

Further, we denote $\boldsymbol{\zeta} = [\boldsymbol{\nu}^T, \boldsymbol{\beta}_0^T, \ldots, \boldsymbol{\beta}_L^T, \boldsymbol{b}_{01}^T, \ldots, \boldsymbol{b}_{LS}^T, \boldsymbol{b}'^T_{011}, \ldots, \boldsymbol{b}'^T_{LPS}]^T$ as a vector of coefficients for the design matrix of observed heterogeneity. With the above notations, we can specify the proposed degradation model in vector form, i.e., $\boldsymbol{y}_i = \boldsymbol{\Omega}_i \boldsymbol{\zeta} + \Lambda_i \boldsymbol{\gamma}_i + \boldsymbol{\varepsilon}_i, i = 1, \ldots, N$.

## B. Parameter estimation under EM framework

### B. 1. Derivation of Eq. (7)

Assume that the measurements of test unit $i$ are normally distributed, the marginal density can then be expressed as $\boldsymbol{y}_i \sim N(\boldsymbol{\Omega}_i \boldsymbol{\zeta}, \Lambda_i \Sigma_\gamma \Lambda_i^T + \sigma_\varepsilon^2 \mathbf{I}_{m_i})$. Given that $\boldsymbol{\gamma}_i$ is known, the conditional distribution becomes $\boldsymbol{y}_i | \boldsymbol{\gamma}_i \sim N(\boldsymbol{\Omega}_i \boldsymbol{\zeta} + \Lambda_i \boldsymbol{\gamma}_i, \sigma_\varepsilon^2 \mathbf{I}_{m_i})$. The latent variables $\boldsymbol{\gamma}_i$ is assumed to be normally distributed, i.e., $\gamma_i \sim N(0, \Sigma_\gamma)$. Based on Bayes rule, the conditional density $p(\gamma_i | y_i, \Theta)$ can be calculated as

$$p(\gamma_i | y_i, \Theta) = \frac{p(y_i | \gamma_i, \Theta) p(\gamma_i | \Theta)}{p(y_i | \Theta)}$$

$$= \frac{C_{11} \frac{1}{\sigma_\varepsilon} |\Sigma_\gamma|^{-\frac{1}{2}} \exp\left[-\frac{1}{2\sigma_\varepsilon^2} \|y_i - \Omega_i \zeta - \Lambda_i \gamma_i\|^2 - \frac{1}{2} \gamma_i^T \Sigma_\gamma^{-1} \gamma_i\right]}{C_{12} |\Lambda_i \Sigma_\gamma \Lambda_i^T + \sigma_\varepsilon^2 I_{m_i}|^{-\frac{1}{2}} \exp\left[-\frac{1}{2} (y_i - \Omega_i \zeta)^T (\Lambda_i \Sigma_\gamma \Lambda_i^T + \sigma_\varepsilon^2 I_{m_i})^{-1} (y_i - \Omega_i \zeta)\right]}$$

where $C_{11}$ and $C_{12}$ are normalizing constants. With the expansion of Euclidean norm, i.e., $\|y_i - \Omega_i \zeta - \Lambda_i \gamma_i\|^2 = \|y_i - \Omega_i \zeta\|^2 + \|\Lambda_i \gamma_i\|^2 - 2(y_i - \Omega_i \zeta)^T \Lambda_i \gamma_i$, the conditional density can be further represented as

$$p(\gamma_i | y_i, \Theta) \propto \frac{|\sigma_\varepsilon^2 \Sigma_\gamma|^{-\frac{1}{2}}}{|\Lambda_i \Sigma_\gamma \Lambda_i^T + \sigma_\varepsilon^2 I_{m_i}|^{-\frac{1}{2}}} \exp[-\frac{1}{2\sigma_\varepsilon^2}((y_i - \Omega_i \zeta)^T (y_i - \Omega_i \zeta) + \gamma_i^T \Lambda_i^T \Lambda_i \gamma_i$$
$$- 2(y_i - \Omega_i \zeta)^T \Lambda_i \gamma_i) - \frac{1}{2} \gamma_i^T \Sigma_\gamma^{-1} \gamma_i + \frac{1}{2}(y_i - \Omega_i \zeta)^T (\Lambda_i \Sigma_\gamma \Lambda_i^T + \sigma_\varepsilon^2 I_{m_i})^{-1} (y_i - \Omega_i \zeta)]$$

$$\propto \left|(\Sigma_\gamma^{-1} + \frac{\Lambda_i^T \Lambda_i}{\sigma_\varepsilon^2})^{-1}\right|^{-\frac{1}{2}} \exp[-\frac{1}{2}(y_i - \Omega_i \zeta)^T (\Lambda_i \Sigma_\gamma \Lambda_i^T + \sigma_\varepsilon^2 I_{m_i})^{-1} (\frac{\Lambda_i \Sigma_\gamma \Lambda_i^T}{\sigma_\varepsilon^2} + I_{m_i})$$
$$\cdot (y_i - \Omega_i \zeta) + \frac{1}{2}(y_i - \Omega_i \zeta)^T (\Lambda_i \Sigma_\gamma \Lambda_i^T + \sigma_\varepsilon^2 I_{m_i})^{-1} (y_i - \Omega_i \zeta) - \frac{1}{2} \gamma_i^T \frac{\Lambda_i^T \Lambda_i}{\sigma_\varepsilon^2} \gamma_i$$
$$+ \frac{1}{\sigma_\varepsilon^2}(y_i - \Omega_i \zeta)^T \Lambda_i \gamma_i - \frac{1}{2} \gamma_i^T \Sigma_\gamma^{-1} \gamma_i]$$

$$\propto \left|(\Sigma_\gamma^{-1} + \frac{\Lambda_i^T \Lambda_i}{\sigma_\varepsilon^2})^{-1}\right|^{-\frac{1}{2}} \exp[-\frac{1}{2}(y_i - \Omega_i \zeta)^T (\Lambda_i \Sigma_\gamma \Lambda_i^T + \sigma_\varepsilon^2 I_{m_i})^{-1} \frac{\Lambda_i \Sigma_\gamma \Lambda_i^T}{\sigma_\varepsilon^2}(y_i - \Omega_i \zeta)$$
$$- \frac{1}{2} \gamma_i^T (\Sigma_\gamma^{-1} + \frac{\Lambda_i^T \Lambda_i}{\sigma_\varepsilon^2}) \gamma_i + \frac{(y_i - \Omega_i \zeta)^T \Lambda_i}{\sigma_\varepsilon^2} \gamma_i]$$

$$\propto |V_i|^{-\frac{1}{2}} \exp\left[-\frac{1}{2}(\gamma_i - \mu_i)^T V_i^{-1} (\gamma_i - \mu_i)\right]$$

where $\mu_i = \frac{1}{\sigma_\varepsilon^2} V_i \Lambda_i^T (y_i - \Omega_i \zeta)$ and $V_i = \left(\Sigma_\gamma^{-1} + \frac{\Lambda_i^T \Lambda_i}{\sigma_\varepsilon^2}\right)^{-1}$. Thus, $\gamma_i | y_i, \Theta$ has gaussian density with mean $\mu_i$ and variance $V_i$. In the E-step of EM estimation framework, we want to compute the expectation of $\gamma_i$ given the observed data and current parameter estimates. With the above derived conditional density, the conditional expectation quantities can then be obtained as

$$E[\gamma_i | D, \Theta] = \mu_i = \frac{1}{\sigma_\varepsilon^2} V_i \Lambda_i^T (y_i - \Omega_i \zeta)$$

$$E[\gamma_i \gamma_i^T | D, \Theta] = V(\gamma_i | D, \Theta) + E[\gamma_i | D, \Theta] (E[\gamma_i | D, \Theta])^T$$
$$= V_i + \mu_i \mu_i^T, \forall i = 1, ..., N$$

Further, the conditional expectation for all test units can be simplified as

$$E[\gamma | D, \Theta] = \mu = \begin{pmatrix} E[\gamma_1 | D, \Theta] \\ E[\gamma_2 | D, \Theta] \\ ... \\ E[\gamma_N | D, \Theta] \end{pmatrix} = (\mu_1^T \mu_2^T ... \mu_N^T)^T$$

## B. 2. Derivation of Eq. (8), Eq. (9) and Eq. (10)

In the M step of EM estimation framework, the parameter estimates $\hat{\Theta}$ can be obtained by maximizing Q function $Q(\Theta, \Theta^{(\tau-1)})$ (as demonstrated in Eq. (6)). Specifically, the parameter estimate $\hat{\zeta}$ at iteration $\tau$ can be obtained by $\hat{\zeta}^{(\tau)} = \arg\max_\zeta Q(\Theta, \Theta^{(\tau-1)})$. This can be further explicitly derived as

$$\frac{\partial Q}{\partial \zeta} = -\frac{1}{2\sigma_\varepsilon^2}\left(\frac{\partial \zeta^T \Omega^T \Omega \zeta}{\partial \zeta} - \frac{\partial \zeta^T \Omega^T y}{\partial \zeta} - \frac{\partial y^T \Omega \zeta}{\partial \zeta} + \frac{\partial E[\gamma^T \mid D, \Theta^{(\tau-1)}]\Lambda^T \Omega \zeta}{\partial \zeta}\right.$$
$$\left. + \frac{\partial \zeta^T \Omega^T \Lambda E[\gamma^T \mid D, \Theta^{(\tau-1)}]}{\partial \zeta}\right) = 0$$

The above equation can be simplified as

$$2\Omega^T \Omega \zeta + 2\Omega^T \Lambda E[\gamma \mid D, \Theta^{(\tau-1)}] - 2\Omega^T y = 0$$

By solving the above equation, we can obtain the parameter estimate as

$$\hat{\zeta}^{(\tau)} = (\Omega^T \Omega)^{-1} \Omega^T (y - \Lambda E[\gamma \mid D, \Theta^{(\tau-1)}]) = (\Omega^T \Omega)^{-1} \Omega^T (y - \Lambda \mu^{(\tau-1)})$$

Similarly, $\hat{\Sigma}_\gamma$ at iteration $\tau$ can be obtained by maximizing Q function, i.e., $\hat{\Sigma}_\gamma^{(\tau)} = \arg\max_{\Sigma_\gamma} Q(\Theta, \Theta^{(\tau-1)})$. This can be achieved by solving the following equation

$$\frac{\partial Q}{\partial \Sigma_\gamma} = -\frac{N}{2}\frac{\partial \log|\Sigma_\gamma|}{\partial \Sigma_\gamma} - \frac{1}{2}\sum_{i=1}^{N}\left(\frac{\partial E[\gamma_i^T \Sigma_\gamma^{-1} \gamma_i \mid D, \Theta^{(\tau-1)}]}{\partial \Sigma_\gamma}\right) = 0$$

By solving the above equation, we can then obtain the parameter estimate as

$$\hat{\Sigma}_\gamma^{(\tau)} = \frac{1}{N}\sum_{i=1}^{N} E[\gamma_i \gamma_i^T \mid D, \Theta^{(\tau-1)}] = \frac{1}{N}\sum_{i=1}^{N}\left(V_i^{(\tau-1)} + \mu_i^{(\tau-1)}\mu_i^{(\tau-1)T}\right)$$
$$= \frac{1}{N}\sum_{i=1}^{N}\left((\hat{\Sigma}_\gamma^{(\tau-1)})^{-1} + \frac{1}{\hat{\sigma}_\varepsilon^{2(\tau-1)}}\Lambda_i^T \Lambda_i)^{-1} + \mu_i^{(\tau-1)}\mu_i^{(\tau-1)T}\right)$$

Based on the updated estimates of $\hat{\zeta}^{(\tau)}$ and $\hat{\Sigma}_\gamma^{(\tau)}$ from above equations, the parameter estimate $\hat{\sigma}_\varepsilon^2$ at iteration $\tau$ can then be obtained by $\hat{\sigma}_\varepsilon^{2(\tau)} = \arg\max_{\sigma_\varepsilon^2} Q(\Theta, \Theta^{(\tau-1)})$. Further, this can be achieved by solving the following equation

$$\frac{\partial Q}{\partial \sigma_\varepsilon^2} = -\frac{\sum_{i=1}^{N} m_i}{2}\frac{\partial \log(\sigma_\varepsilon^2)}{\partial \sigma_\varepsilon^2} - \frac{\partial \frac{1}{2\sigma_\varepsilon^2}}{\partial \sigma_\varepsilon^2}(\|y - \Omega\hat{\zeta}^{(\tau)}\|^2 + \sum_{i=1}^{N}\text{Tr}(\Lambda_i^T \Lambda_i E[\gamma_i \gamma_i^T \mid D, \Theta^{(\tau-1)}])$$
$$- 2(y - \Omega\hat{\zeta}^{(\tau)})^T \Lambda E[\gamma \mid D, \Theta^{(\tau-1)}]) = 0$$

By solving the above equation, the closed form of parameter estimate then becomes

$$\hat{\sigma}_\varepsilon^{2(\tau)} = \frac{1}{\sum_{i=1}^{N} m_i}(\|y - \Omega\hat{\zeta}^{(\tau)}\|^2 - 2\sum_{i=1}^{N}(y_i - \Omega_i\hat{\zeta}^{(\tau)})^T \Lambda_i E[\gamma_i \mid D, \Theta^{(\tau-1)}]$$
$$+ \sum_{i=1}^{N}\text{Tr}(\Lambda_i^T \Lambda_i E[\gamma_i \gamma_i^T \mid D, \Theta^{(\tau-1)}])$$
$$= \frac{1}{\sum_{i=1}^{N} m_i}(\|y - \Omega\hat{\zeta}^{(\tau)}\|^2 - 2\sum_{i=1}^{N}(y_i - \Omega_i\hat{\zeta}^{(\tau)})^T \Lambda_i \mu_i^{(\tau-1)}$$
$$+ \sum_{i=1}^{N}\text{Tr}(\Lambda_i^T \Lambda_i ((\hat{\Sigma}_\gamma^{(\tau)})^{-1} + \frac{1}{\hat{\sigma}_\varepsilon^{2(\tau-1)}}\Lambda_i^T \Lambda_i)^{-1} + \mu_i^{(\tau-1)}\mu_i^{(\tau-1)T})))$$

Based on the above derived closed forms of parameter estimates, the estimation procedure will iteratively update the parameters through expectation step and maximization step.

### Acknowledgments

W. Cai thankfully acknowledges the financial support by the National Science Foun- dation under grant CMMI-1855651. The CuNiSn bronze specimens are provided by Brush-Wellman Inc.